\documentclass[runningheads,hidelinks]{llncs}
\usepackage{booktabs}
\usepackage{multirow}
\usepackage{tabularx} 
\usepackage[T1]{fontenc}
\usepackage{subcaption} %
\usepackage{graphicx}
\usepackage{hyperref}
\usepackage{color}

\begin{document}
\title{Xenophobic Events vs. Refugee Population – Using GDELT to Identify Countries with Disproportionate Coverage}
\titlerunning{Xenophobic Events vs. Refugee Population}
%
\author{Himarsha R. Jayanetti\inst{1}\orcidID{0000-1111-2222-3333} \and
Erika Frydenlund\inst{2}\orcidID{0000-0002-7694-7845} \and
Michele C. Weigle\inst{1}\orcidID{0000-0002-2787-7166}
}
\authorrunning{H. R. Jayanetti et al.}
%
\institute{Old Dominion University, Norfolk, Virginia, USA \\
\email{\{hjaya002\}@odu.edu}
\email{\{mweigle\}@cs.odu.edu} \and
Virginia Modeling, Analysis and Simulation Center, Old Dominion University, Suffolk, Virginia, USA \\
\email{\{efrydenl\}@odu.edu} 
}
\maketitle 
\begin{abstract}

In this preliminary study, we used the Global Database of Events, Language, and Tone (GDELT) database to examine xenophobic events reported in the media during 2022. We collected a dataset of 2,778 unique events and created a choropleth map illustrating the frequency of events scaled by the refugee population's proportion in each host country. We identified the top 10 countries with the highest scaled event frequencies among those with more than 50,000 refugees. Contrary to the belief that hosting a significant number of forced migrants results in higher xenophobic incidents, our findings indicate a potential connection to political factors. We also categorized the 20 root event codes in the CAMEO event data as either ``Direct'' or ``Indirect''. Almost 90\% of the events related to refugees in 2022 were classified as ``Indirect''.

\end{abstract}

\section{Introduction}
\label{sec:intro}
People move around the world in pursuit of better opportunities or to flee conflicts and natural disasters. There are 281 million international migrants, or one in every 30 people worldwide~\cite{436}, and more than 82 million of them have been forcibly displaced~\cite{unhcrfaag}. These migrants make an effort to coexist in their host communities. However, widespread xenophobic and racist violence makes it difficult to uphold societal order and provide equal access to opportunities, resources, and even human dignity. Hence, it is imperative to study such xenophobic incidents and examine the underlying factors contributing to hostile behavior towards refugees in order to fight xenophobia. 

In our preliminary study, we use a massive and regularly updated dataset of online, TV, and news reporting from Global Data on Events, Location, and Tone (GDELT)\footnote{\url{https://www.gdeltproject.org/}} to look for ways to characterize xenophobic events. By collecting data from 2022, we were able to analyze and derive meaningful insights from the dataset within that one-year period. 

Despite GDELT's richness as a data source, harnessing it for very specific tasks like tracking xenophobic events presents challenges. Hoffmann et al.~\cite{hoffmann_2022} pointed out the challenges of utilizing big data news sources, such as the need for a deep understanding of the database's complexity and a significant time investment in data analysis. 
Hammond and Weidmann~\cite{Hammond_2014} evaluated the applicability of GDELT data studying political violence. The results emphasize the need for caution when using GDELT for geospatial studies through a comparison with two hand-coded conflict event datasets. 
Hence, conducting exploratory research using a short period, such as a single year (2022), is crucial for better comprehension of the challenges involved.

In this paper, we are addressing two key research questions (RQ1 and RQ2):
\begin{enumerate}
    \item RQ1: Can we identify countries with an unexpectedly high number of refugee-related events reported in the news media relative to their refugee populations?
    \item RQ2: Can the 20 root event codes be broadly categorized into ``Direct'' and ``Indirect'', and how prevalent are ``Indirect'' actions compared to ``Direct'' actions within our dataset?
\end{enumerate}

Our findings challenge the notion that hosting a large number of forced migrants directly correlates with an increase in xenophobic incidents, suggesting a complex interplay of factors beyond refugee populations and highlighting the potential influence of political dynamics. Moreover, we were able to group the 20 root event codes into ``Direct'' and ``Indirect'' categories and found that a significantly higher proportion were classified as ``Indirect'' actions.

\section{Background and Related Work}
\label{sec:background}
GDELT is a prime example of big data, with billions of records spanning from 1979 to the present. It uses the Conflict and Mediation Event Observations (CAMEO) taxonomy, a framework for coding event-related data, to automatically code data for use in research~\cite{gerner2002conflict}. 
Researchers  have used GDELT data for a variety of studies, such as studying the effects of civil unrest, complexity in terms of political activities, and capturing peace through the Global Peace Index (GPI)~\cite{fang2016identifying,voukelatou2022understanding,yonamine2013nuanced}. Vargo et al. studied the power of fake news from 2014 to 2016 in online news media using GDELT~\cite{vargo2018agenda}. Their research revealed that although the prevalence of fake news has risen, these websites do not possess undue influence. Other researchers have used social media platforms as well as newspaper articles in opinion mining about a range of topics from online education to industrial production~\cite{fu2020opinion,tilly2021macroeconomic}. 

Various studies have examined forced migration and policy implications in countries that support migration~\cite{frydenlund2019mobility,frydenlund2022opportunities,frydenlund2019characterizing}. Napierala et al. summarize how big data, like GDELT, have been used to help predict refugee flows and humanitarian situations~\cite{early_warning}. Yesilbas et al. utilized GDELT to build a large dataset of global news to study the tone, volume, and topics of media coverage of refugees~\cite{yesilbas2021analysis}. They found that a negative tone occurred both because of the anti-migrant sentiment as well as sorrow and empathy for the refugees. Weisser used GDELT to characterize the actors (e.g., nonprofits, businesses, and government) that drive the conversation and support for refugees and migrants~\cite{near_real_time}. 

Our research shares similarities with previous studies in that we aim to analyze events related to refugee and migrant communities. Our approach differs from previous analyses and datasets in that our final objective is to develop a real-time monitoring system for xenophobic events utilizing GDELT data to identify potential hotspots of violence to predict potential escalation. While combating xenophobia is within the mandates of international organizations such as the United Nations High Commissioner for Refugees (UNHCR) and the International Organization for Migration (IOM), there is no worldwide tool for tracking these events. The Internal Displacement Monitoring Centre\footnote{\url{https://www.internal-displacement.org}} has designed a widely-adopted hand-coded data synthesis tool to monitor internal displacement caused by conflict and natural disasters. ACLED\footnote{\url{https://acleddata.com/}} similarly tracks protest and violence across the world. Xenowatch\footnote{\url{https://www.xenowatch.ac.za}} is an online heatmap using data that researchers have hand-coded from user-submitted news articles of xenophobic events in South Africa. These tools have had immeasurable impacts on research and decision-making, but no such tool exists on the global scale for xenophobic events. The ultimate aim of this study is to construct such a tool; however, this paper discusses our preliminary exploratory analysis of 2022 global news data automatically coded by GDELT for events involving migrants.


\section{Methodology}
\label{sec:method}




\subsection{Understanding the GDELT Database}
\label{subsec:understanding_data}
In our study, we are using the GDELT 2.0 database, which is updated every 15 minutes and translates articles from around the world from 65 different languages into English~\cite{gdelt2}. In our exploration of the use of GDELT, we identified three key tables in the database:

\begin{enumerate}
    \item \textbf{Event}: Contains data about events happening globally. Each row represents a single event, coded with information such as an event identification number (\texttt{GLOBALEVENTID}), actors (\texttt{Actor1Code}, \texttt{Actor2Code}), and location (\texttt{Actor1CountryCode}, \texttt{Actor2CountryCode}).
    
    \item \textbf{Event Mentions}: Contains a row for each mention of the event in a news article or other source. Each mention is coded with its respective \texttt{GLOBALEVENTID} and information about the tone of the mention (positive or negative). The \texttt{GLOBALEVENTID} corresponds to the \texttt{GLOBALEVENTID} from the Event Table, and the external identifier \texttt{MentionIdentifier} for the source document uniquely identifies the document. 
    
    \item \textbf{Global Knowledge Graph (GKG)}~\cite{gdeltgkg}: Connects data from various sources to form an interconnected network that encapsulates events around the world, their corresponding contexts, associated actors, and the overall sentiment of media coverage. The \texttt{DocumentIdentifier} field corresponds to the \texttt{MentionIdentifier} in the Event Mentions table. This GKG table comprises the results obtained from the Global Content Analysis Measures (GCAM) system, which employs multiple  content analysis tools to capture over 2,230 latent dimensions for each news article monitored by GDELT~\cite{gdeltgcam}.
    
\end{enumerate}




\subsection{Data Collection Methods and Criteria}
\label{subsec:data_collection}

We used the year 2022 for our analysis after considering several factors. Firstly, we aimed to focus on a recent timeframe with the intention of building upon our understanding and awareness of recent events. We intentionally avoided selecting the year 2021 due to COVID-19 disruptions, including the peak of the pandemic and related restrictions. We avoided recent years due to challenges in accounting for changes in migration patterns during the pandemic due to lockdown measures. By focusing on 2022, we achieved a balance between recentness and mitigating potential confounding factors that could affect our findings. 


To capture xenophobic events in the media, we identified 8 main GKG themes~\cite{gkg_themes} that are related to negative actions against immigrants, which we refer to collectively as ``GKG\_REF'':
    
    \begin{itemize}
        \item \texttt{DISCRIMINATION\_IMMIGRATION\_XENOPHOBIA}
        \item \texttt{DISCRIMINATION\_IMMIGRATION\_ANTIIMMIGRANTS}
        \item \texttt{DISCRIMINATION\_IMMIGRATION\_OPPOSED\_TO\_IMMIGRANTS}
        \item \texttt{DISCRIMINATION\_IMMIGRATION\_AGAINST\_IMMIGRANTS}
        \item \texttt{DISCRIMINATION\_IMMIGRATION\_ATTACKS\_ON\_IMMIGRANTS}
        \item \texttt{DISCRIMINATION\_IMMIGRATION\_ATTACKS\_AGAINST\_IMMIGRANTS}
        \item \texttt{DISCRIMINATION\_IMMIGRATION\_XENOPHOBE}
        \item \texttt{DISCRIMINATION\_IMMIGRATION\_XENOPHOBES} 
    \end{itemize}

\begin{figure}
{
\centering
\includegraphics[width=\textwidth]{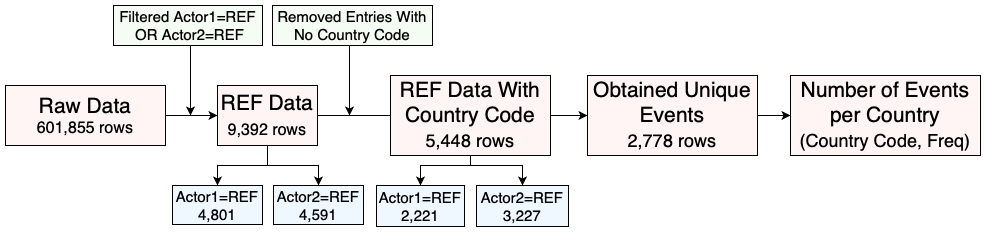}
\caption{Data analysis and filtering process}
\label{fig:method}
}
\end{figure}

Figure~\ref{fig:method} depicts the steps of our data collection and filtering process. We used Google BigQuery\footnote{\url{https://cloud.google.com/bigquery}} to connect to the GDELT databases and collected data for 2022 having ``GKG\_REF'' themes (using the query filter \texttt{V2Themes like DISCRIMINATION\_IMMIGRATION}). This initial dataset comprised 601,855 records from the combined Event, Event Mentions, and GKG tables. After filtering for Actor1Code or Actor2Code corresponding to REF (actor code for "Refugee"), we obtained 9,392 records. We further filtered out entries without a country code, resulting in 5,448 records. Among these, we identified 2,778 unique events, which were saved as a CSV file with the country code and event frequency.




Our goal is to highlight those countries that have a greater number of refugee-related events reported in news media than would be expected based on the relative size of their refugee populations. This might point to countries where refugees have become a contentious political issue. To do this, we first calculated the proportion of refugees per host country by dividing the number of refugees by the country's total population (RT). We obtained the population of displaced people (refugees, asylum seekers, and others in refugee-like situations) for each country of asylum (as of mid-2022) from UNHCR\footnote{\url{https://www.unhcr.org/refugee-statistics/download/?url=c1lN2c}} and the total country population for 2022 from the US Census Bureau.\footnote{\url{https://www.census.gov/programs-surveys/international-programs/data.html}} Then, we divided the number of relevant GDELT events per country by that country's proportion of refugees. For instance, in 2022, Turkey and Germany had similar total populations ($\sim$ 80 million) and a similar number of relevant GDELT events ($\sim 100$). But, since Germany's proportion of refugees to the total population ($\sim 3\%$) was lower than Turkey's ($\sim 5\%$), Germany was given a higher scaled frequency score.


%

\section{Results}
\label{sec:results}

\subsection{Scaled Frequency of Events Per Country}
\label{subsec:eventfreq}

\begin{figure}
{
\centering
\fbox{\includegraphics[width=0.97\textwidth, trim=40 120 50 50, clip]{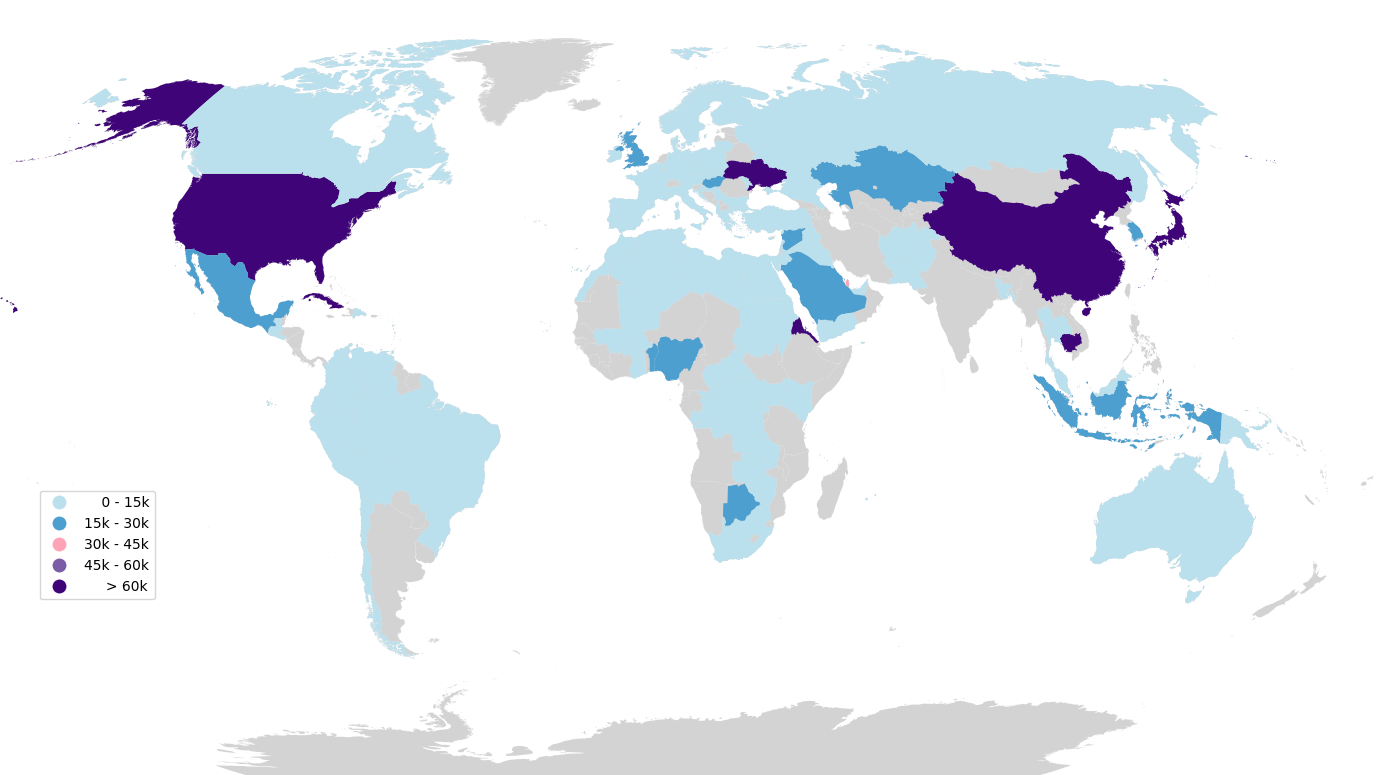}}
\caption{Choropleth map illustrating the scaled frequency of unique events}
\label{fig:map_freq_all}
}
\end{figure}

Figure~\ref{fig:map_freq_all}  is a choropleth map showing the scaled frequency of the number of unique events in the data. 
Gray areas indicate countries without available data. Dark purple represents countries with the highest scaled event frequencies, meaning they had more relevant events than expected based on their proportion of refugees. Light blue represents countries with the lowest scaled event frequencies.

Of note in Figure~\ref{fig:map_freq_all} is that three countries -- the USA, China, and Ukraine -- produce the largest frequency of reported events of discrimination/violence against migrants relative to the volume of refugees hosted in their countries. This may be explained by one of the biases in news coverage of GDELT, where a significant number of media sources originate from the US and China. Additionally, Ukraine is the third largest and also most recent refugee situation in the world. The recency of this event may explain its prevalence in news coverage.   


Table~\ref{tab:top10_scaled} shows the top 10 countries based on the scaled frequency of unique events. When reporting the top 10 countries, we established a threshold of hosting a minimum of 50,000 refugees and other displaced people. This decision was made to prevent a significant decrease in the RT value, which in turn would greatly increase the scaled frequency.


According to the UNHCR, 36\% of all displaced people are hosted in just five countries: Turkey, Colombia, Germany, Pakistan, and Uganda. Notably, none of these  are in the top 10 by scaled frequency.
Counter to the idea that hosting a large number of forced migrants increases xenophobic events, Nigeria (NGA) is in the top three countries, but the number of displaced people hosted is quite low. Rwanda (RWA) additionally has a disproportionate number of xenophobic events coverage given how small the national population is compared to others in the top 10 list. Together, these entries suggest that xenophobic events coverage is related to more than just the burden of hosting large numbers of refugees, but indicates a potential association with political dynamics.






\begin{table}
\caption{Top 10 countries based on the scaled frequency of unique events}\label{tab:top10_scaled}
\centering
\resizebox{\textwidth}{!}{%
\begin{tabular}{| r | r | r | r | r | r |}
\hline 

\multicolumn{1}{|c|}{Country}  & \multicolumn{1}{|c|}{Event} & \multicolumn{1}{c|}{Refugee} & \multicolumn{1}{c|}{Total}  & \multicolumn{1}{c|}{Refugee/Total}  &   \\ 
\multicolumn{1}{|c|}{Code} & \multicolumn{1}{c|}{Frequency} & \multicolumn{1}{c|}{Population} & \multicolumn{1}{c|}{Population} & \multicolumn{1}{c|}{Population} & \multicolumn{1}{c|}{Scaled Frequency} \\ 
\multicolumn{1}{|c|}{(CC)} & \multicolumn{1}{c|}{(F)} & \multicolumn{1}{c|}{(RP)} & \multicolumn{1}{c|}{(TP)} & \multicolumn{1}{c|}{(RT)} & \multicolumn{1}{c|}{(F/RT)} \\
 \hline 

USA	&	354 	&	1,787,504	&	337,341,954	&	0.0053	&	66,807.71	\\
GBR	&	158	&	359,311	&	67,791,400	&	0.0053	&	29,809.95	\\
NGA	&	9	&	84,302	&	225,082,083	&	0.0004	&	24,029.55	\\
MEX	&	59	&	498,213	&	129,150,971	&	0.0039	&	15,294.48	\\
ITA	&	62	&	354,414	&	61,095,551	&	0.0058	&	10,687.85	\\
CAN	&	28	&	126,499	&	38,232,593	&	0.0033	&	8,462.62	\\
RUS	&	83	&	1,463,050	&	142,021,981	&	0.0103	&	8,057.02	\\
ZAF	&	32	&	240,077	&	57,516,665	&	0.0042	&	7,666.43	\\
RWA	&	73	&	128,056	&	13,173,730	&	0.0097	&	7,509.86	\\
BRA	&	17	&	562,577	&	217,240,060	&	0.0026	&	6,564.58	\\
\hline
\end{tabular}  
}
\end{table}

\subsection{Event Root Code Classification}
\label{subsec:eventrootcode}

To explore this further, we characterized the CAMEO event codes~\cite{cameo_codebook}, which broadly label the news articles using action words describing the news coverage. 
These codes are organized into 20 root event codes, numbered from 01 to 20. We categorized these into two broad categories, ``Direct'' and ``Indirect'' actions. By indirect, we look at actions related to diplomacy and ``talking'' about an issue, and direct actions are those that require physical actions, such as ``demanding'' or even ``fighting'' or ``assaulting''. Table~\ref{tab:eventrootcode} shows the classification of the 20 root codes into our two predefined categories. Out of the total 2,778 events, we classified 360 (13\%) as ``Direct'' and 2,418 (87\%)  as ``Indirect''.


\begin{table*}
\centering
\caption{Cameo event root codes, its description, and our categorization}
\label{tab:eventrootcode}
\begin{tabular}{c l  c l}
\hline
\multicolumn{2}{c}{\textbf{Indirect}} & \multicolumn{2}{c}{\textbf{Direct}} \\
\hline
\textbf{Event} & \textbf{} & \textbf{Event} & \textbf{} \\

\textbf{RootCode} & \multicolumn{1}{c}{\textbf{Description}} & \textbf{RootCode} & \multicolumn{1}{c}{\textbf{Description}} \\

\hline

01	&	Make Public Statement	& 09	&	Investigate	\\
02	&	Appeal	& 10	&	Demand	\\
03	&	Express Intent to Cooperate	 & 13	&	Threaten	\\
04	&	Consult & 14	&	Protest	\\
05	&	Engage in Diplomatic Cooperation	& 15	&	Exhibit Force Posture	\\
06	&	Engage in Material Cooperation & 16	&	Reduce Relations	\\
07	&	Provide Aid	 & 17	&	Coerce	\\
08	&	Yield	& 18	&	Assault	\\
11	&	Disapprove	& 19	&	Fight	\\
12	&	Reject	& 20	&	Use Unconventional Mass		\\

\hline 

\end{tabular}
\end{table*}





Table~\ref{tab:top10_rootcode} presents a breakdown of data by country code of the top 10 countries, specifically showcasing the number of events for the ``Direct'' and ``Indirect'' categories, as well as the total frequency of events. It captures the idea that much of the media coverage of discrimination and/or violence against refugees and other displaced people is largely ``all talk and no action''. Assuming that an actual xenophobic action would fall under a ``direct'' event (which contains assault, fight, protest, and threaten actions), 2022 saw very few events in locations that have a large relative frequency of event reporting. Nigeria, for instance, is in the top 10 countries reporting the most on discrimination and/or violence against refugees but had zero ``direct'' events. 

\begin{table}[h]
\caption{Number of ``Indirect'' and ``Direct'' events for the top 10 countries}\label{tab:classification}
\centering
\label{tab:top10_rootcode}
\begin{tabular}{| c | r | r | r |}
\hline 
\multicolumn{1}{|c|}{Country Code}  &  \multicolumn{1}{c|}{Indirect} &  \multicolumn{1}{c|}{Direct} & \multicolumn{1}{c|}{Total}   \\ 
\hline 

USA	&	298 (84\%) 	&	56 (16\%)	&	354	\\
GBR	&	142 (90\%)	&	16 (10\%)	&	158	\\
NGA	&	9 (100\%)	&	0 (0\%)	&	9	\\
MEX	&	52 (88\%)	&	7 (12\%)	&	59	\\
ITA	&	49 (79\%)	&	13 (21\%)	&	62	\\
CAN	&	25 (89\%)	&	3 (11\%)	&	28	\\
RUS	&	73 (88\%)	&	10 (12\%)	&	83	\\
ZAF	&	31 (97\%)	&	1 (3\%)	&	32	\\
RWA	&	60 (82\%)	&	13 (18\%)	&	73	\\
BRA	&	9 (53\%)	&	8 (47\%)	&	17	\\

\hline
\end{tabular}  
\end{table}

Further, potentially qualitative deep-analysis of newspaper articles is required to understand whether indirect actions are a predictor of future xenophobic events. What is clear is that actual events are disproportionately outnumbered by news coverage that talks about doing something. While large datasets like GDELT can provide evidence of these trends, further review of samples of the actual news articles is required to understand how these CAMEO codes may be refined and tested as predictors of upticks in direct xenophobic events.



\section{Future Work}
\label{sec:futurework}

When considering future work, we need to address some limitations in our study. Firstly, we only analyzed events with location information in the dataset. Therefore, it would be valuable to investigate events without location information and explore their characteristics and implications. Additionally, our study focused on a single year, specifically 2022. To gain a better understanding of patterns and phenomena, it is important to extend the study to a wider time range and assess the robustness and consistency of the identified findings. Furthermore, we can use the results of our study to develop improved interactive visualizations for monitoring xenophobic violence against refugees and migrants. 




\section{Conclusion}
\label{sec:conclusions}
In this study, we used the GDELT database to explore and characterize xenophobic events reported in the media in 2022. Our goal was to highlight those countries that had a greater number of refugee-related events reported in news media than would be expected based on the relative size of their refugee populations (RQ1). We identified 8 GKG themes, collectively referred to as ``GKG\_REF’’ present in the GDELT dataset. Our dataset is comprised of 2,778 unique events involving ``REF’’ actors (actor code for refugees in GDELT) having ``GKG\_REF’’ themes. The results of our analysis revealed countries with higher scaled event frequencies, indicating a greater number of xenophobic events reported in the media relative to their refugee populations. Notably, the United States, China, and Ukraine emerged as countries with a significant frequency of reported events despite variations in their refugee populations.  We reported the top 10 countries with the highest scaled event frequencies  (of those having more than 50,000 refugees). Counter to the idea that hosting a large number of forced migrants leads to an increase in xenophobic incidents, our findings showed otherwise. Our results suggested that xenophobic events coverage is related to more than just the burden of hosting large numbers of refugees, but indicating a potential association with political dynamics. 
Further, we grouped the 20 root event codes in CAMEO event categories into two primary categories: ``Direct'' and ``Indirect'' actions (RQ2). Out of the total 2,778 events, 13\% (360) were classified as ``Direct'', while 87\% (2,418) were categorized as ``Indirect'', suggesting a greater prevalence of indirect actions like diplomacy and discourse compared to direct actions such as demanding or assaulting. If we consider that xenophobic actions are typically categorized as ``direct’’ events, which include actions like assault, fight, protest, and threaten, there were only a few occurrences in 2022 in the top 10 countries compared to ``indirect’’ events. 
By discussing the challenges associated with using extensive datasets like GDELT for specific tasks, we emphasized the significance of conducting focused exploratory research within a specific timeframe (such as a single year – 2022), to gain a deeper understanding of these challenges.

\subsubsection{Acknowledgements}
\label{sec:acknowledgments}
This research was funded under the project ``Data Science for Social Good: Mining and Visualizing Worldwide News to Monitor Xenophobic Violence'', through the 2022-2023 ODU Data Science Seed Funding Program. 

\bibliographystyle{splncs04}
\bibliography{refs}

\begin{thebibliography}{10}
\providecommand{\url}[1]{\texttt{#1}}
\providecommand{\urlprefix}{URL }
\providecommand{\doi}[1]{https://doi.org/#1}

\bibitem{fang2016identifying}
Fang, P., Gao, J., Fan, F., Yang, L.: Identifying political “hot” spots
  through massive media data analysis. In: Proceedings of the 9th International
  Conference on Social, Cultural, and Behavioral Modeling (SBP-BRiMS). pp.
  282--290 (2016). \doi{10.1007/978-3-319-39931-7_27}

\bibitem{frydenlund2019mobility}
Frydenlund, E., Jones, E.C., Padilla, J.J.: Mobility in crisis: An agent-based
  model of refugees’ flight to safety. Human Simulation: Perspectives,
  Insights, and Applications pp. 191--208 (2019).
  \doi{10.1007/978-3-030-17090-5_11}

\bibitem{frydenlund2022opportunities}
Frydenlund, E., Padilla, J.J.: Opportunities and challenges of using
  computer-based simulation in migration and displacement research: A focus on
  {Lesbos, Greece}. In: Bradley, M., Milner, J. (eds.) Documenting
  Displacement: Questioning Methodological Boundaries in Forced Migration
  Research, chap.~11, pp. 279--308. McGill-Queen’s University Press (2022)

\bibitem{frydenlund2019characterizing}
Frydenlund, E., Yilmaz~{\c{S}}ener, M., Gore, R., Boshuijzen-van Burken, C.,
  Bozdag, E., De~Kock, C.: Characterizing the mobile phone use patterns of
  refugee-hosting provinces in {Turkey}. In: Salah, A., Pentland, A., Lepri,
  B., Letouzé, E. (eds.) Guide to Mobile Data Analytics in Refugee Scenarios,
  pp. 417--431 (2019). \doi{10.1007/978-3-030-12554-7_21}

\bibitem{fu2020opinion}
Fu, Z., Yan, M., Meng, C., Wang, W., Hu, X., Li, Y., Wang, J., He, Z., Wang,
  Z.: Opinion mining about online education basing on {GDELT} and {Twitter}
  data. In: Proceedings of the IEEE/WIC/ACM International Joint Conference on
  Web Intelligence and Intelligent Agent Technology (WI-IAT). pp. 741--745
  (2020). \doi{10.1109/WIIAT50758.2020.00114}

\bibitem{gerner2002conflict}
Gerner, D.J., Schrodt, P.A., Yilmaz, O., Abu-Jabr, R.: Conflict and mediation
  event observations {(CAMEO)}: A new event data framework for the analysis of
  foreign policy interactions. In: Proceedings of the Annual Meeting of the
  International Studies Association (2002)

\bibitem{Hammond_2014}
Hammond, J., Weidmann, N.B.: Using machine-coded event data for the micro-level
  study of political violence. Research \& Politics  \textbf{1}(2),
  2053168014539924 (2014). \doi{10.1177/2053168014539924},
  \url{https://doi.org/10.1177/2053168014539924}

\bibitem{hoffmann_2022}
Hoffmann, M., Santos, F.G., Neumayer, C., Mercea, D.: Lifting the veil on the
  use of big data news repositories: A documentation and critical discussion of
  a protest event analysis. Communication Methods and Measures  \textbf{16}(4),
   283--302 (2022). \doi{10.1080/19312458.2022.2128099},
  \url{https://doi.org/10.1080/19312458.2022.2128099}

\bibitem{436}
McAuliffe, M., Triandafyllidou, A. (eds.): World Migration Report 2022.
  International Organization for Migration (2021)

\bibitem{early_warning}
Napierała, J., Hilton, J., Forster, J.J., Carammia, M., Bijak, J.: Toward an
  early warning system for monitoring asylum-related migration flows in
  {Europe}. International Migration Review  \textbf{56}(1),  33--62 (2022).
  \doi{10.1177/01979183211035736}

\bibitem{cameo_codebook}
{Philip A. Schrodt}: {Conflict and Mediation Event Observations Event and Actor
  Codebook (CAMEO)}.
  \url{http://data.gdeltproject.org/documentation/CAMEO.Manual.1.1b3.pdf}
  (2012)

\bibitem{gdelt2}
{The GDELT Project}: {GDELT 2.0: Our Global World in Realtime}.
  \url{https://blog.gdeltproject.org/gdelt-2-0-our-global-world-in-realtime/}
  (2015)

\bibitem{gkg_themes}
{The GDELT Project}: {New November 2021 GKG 2.0 Themes Lookup}.
  \url{https://blog.gdeltproject.org/new-november-2021-gkg-2-0-themes-lookup/}
  (2021)

\bibitem{gdeltgcam}
{The GDELT Project, GCAM}: {Introducing the Global Content Analysis Measures
  (GCAM)}.
  \url{https://blog.gdeltproject.org/introducing-the-global-content-analysis-measures-gcam/}
  (2014)

\bibitem{gdeltgkg}
{The GDELT Project, GKG}: {Introducing GKG 2.0 – The Next Generation of the
  GDELT Global Knowledge Graph}.
  \url{https://blog.gdeltproject.org/introducing-gkg-2-0-the-next-generation-of-the-gdelt-global-knowledge-graph/}
  (2014)

\bibitem{tilly2021macroeconomic}
Tilly, S., Ebner, M., Livan, G.: Macroeconomic forecasting through news,
  emotions and narrative. Expert Systems with Applications  \textbf{175},
  114760 (2021). \doi{10.1016/j.eswa.2021.114760}

\bibitem{unhcrfaag}
{UNHCR}: {Figures at a Glance}.
  \url{https://www.unhcr.org/en-us/figures-at-a-glance.html} (Jun 2022)

\bibitem{vargo2018agenda}
Vargo, C.J., Guo, L., Amazeen, M.A.: The agenda-setting power of fake news: A
  big data analysis of the online media landscape from 2014 to 2016. New Media
  \& Society  \textbf{20}(5),  2028--2049 (2018).
  \doi{10.1177/1461444817712086}

\bibitem{voukelatou2022understanding}
Voukelatou, V., Miliou, I., Giannotti, F., Pappalardo, L.: Understanding peace
  through the world news. EPJ Data Science  \textbf{11}(2) (2022).
  \doi{10.1140/epjds/s13688-022-00315-z}

\bibitem{near_real_time}
Weisser, R.A.: {A near-real-time analysis of societal responses to {Ukrainian}
  refugee migration in {Europe}}. International Migration  (Nov 2022).
  \doi{10.1111/imig.13071}

\bibitem{yesilbas2021analysis}
Yesilbas, V., Padilla, J.J., Frydenlund, E.: An analysis of global news
  coverage of refugees using a big data approach. In: Proceedings of the 14th
  International Conference on Social, Cultural, and Behavioral Modeling
  (SBP-BRiMS). pp. 111--120 (2021). \doi{10.1007/978-3-030-80387-2_11}

\bibitem{yonamine2013nuanced}
Yonamine, J.E.: A nuanced study of political conflict using the {Global
  Datasets of Events Location and Tone (GDELT)} dataset. Ph.D. thesis, The
  Pennsylvania State University (Aug 2013)

\end{thebibliography}

\end{document}